\documentclass[superscriptaddress,prc,preprint,showpacs,showkeys,floatfix]{revtex4}
\usepackage[dvips]{graphicx}
\begin{document}
\title{Universal expression for the lowest excitation energy \\
of natural parity even multipole states}
\author{Doohwan \surname{Kim}}
\affiliation{ Nuclear Data Evaluation Lab, Korea Atomic Energy
Research Institute, Daejeon 305-353, Korea}
\author{Eunja \surname{Ha}}
\affiliation{Department of Physics, Inha University, Incheon
402-751, Korea}
\author{Dongwoo \surname{Cha}}
\email{dcha@inha.ac.kr}
\
\thanks{Fax: +82-32-866-2452}
\affiliation{Department of Physics, Inha University, Incheon
402-751, Korea}
\date{August 30, 2007}

\begin{abstract}
We present a new expression for the energy of the lowest collective states in even-even nuclei throughout the entire periodic table. Our empirical formula holds universally for all of the natural parity even multipole states and describes the overall trends. This formula depends only on the mass number and the valence nucleon numbers with six parameters. The parameters are determined unambiguously from the data for each multipole state. We discuss the validity of our empirical formula by comparing our results with those of other studies and also by estimating the average and the dispersion of the logarithmic errors of the calculated excitation energies with respect to the measured ones.
\end{abstract}

\pacs{21.10.Re, 23.20.Lv}

\keywords{Empirical formula; Lowest electric state excitation energies; Valence nucleon numbers}

\maketitle

\begin{figure}[h]
\centering
\includegraphics[width=14.0cm,angle=0]{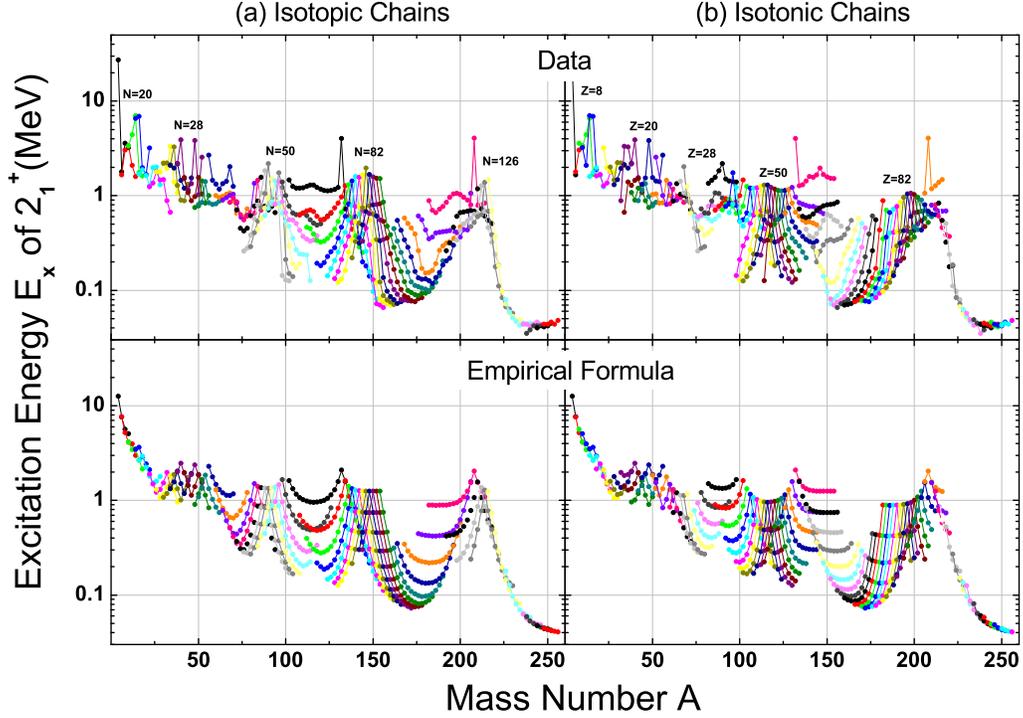}
\caption{The excitation energies of the first $2^+$ states in even-even nuclei. The data points are connected by solid lines along the isotopic chains in (a) and along the isotonic chains in (b). The upper part shows the measured excitation energies while the lower part shows those calculated by our six parameter empirical formula given by Eq.\,(\ref{GE}). The measured excitation energies are quoted from the compilation in Raman {\it et al}. \cite{Raman1}. In fact, the upper part of (a) is a duplicate of Fig.\,I(a) in Ref. \cite{Raman1}.}
\label{fig-1}
\end{figure}

In a previous publication \cite{Ha}, we reported empirical findings
of a simple formula which could reproduce the excitation energy $E_x
(2_1^+)$ for the first $2^+$ states in even-even nuclei. The idea for the particular structure of our empirical formula was first envisaged by inspecting Figs.\,I(a), II(a), and III(a) of Ref.\,\cite{Raman1} where the measured excitation energies of the first $2^+$ states in 557 even-even nuclei were displayed as function of the mass number $A$, the proton number $Z$, and the neutron number $N$, respectively. One of these figures, which is duplicated in the upper part of Fig.\,\ref{fig-1}(a), shows the first $2^+$ excitation energies connected along the isotopic chains. In the upper part of Fig.\,\ref{fig-1}(b), we draw the same graph again but by reconnecting the data points along the isotonic chains. From a given isotopic chain in Fig.\,\ref{fig-1}(a), we can easily recognize that $E_x (2_1^+)$ is minimum in the mid-shell nucleus and that $E_x (2_1^+)$ becomes larger when the neutron number increases or decreases from the mid-shell. Furthermore, as indicated by the numbers in the upper part of Fig.\,\ref{fig-1}(a), the neutron number $N$ of the nucleus, for which $E_x (2_1^+)$ is maximum along a given isotopic chain, is one of the neutron magic numbers 20, 28, 50, 82, and 126. We can also observe similar features in the upper part of Fig.\,\ref{fig-1}(b). For a given isotonic chain, $E_x (2_1^+)$ is minimum in the mid-shell nucleus and $E_x (2_1^+)$ becomes larger when the proton number increases or decreases from mid-shell. The proton number $Z$ of the nucleus, for which $E_x (2_1^+)$ is maximum along a given isotonic chain, is again one of the proton magic numbers 8, 20, 28, 50, and 82.

The above behavior of $E_x (2_1^+)$ can be best represented by employing the valence nucleon numbers $N_p$ and $N_n$. The valence proton (neutron) number $N_p\,(N_n )$ is defined as the number of proton (neutron) particles above the highest filled major shell or  the number of proton (neutron) holes if the Fermi level is beyond the mid-shell within the highest proton (neutron) major shell. Since the valence nucleon number is maximum at the mid-shell nucleus and zero at the top nucleus of a major shell, we can express the excitation energy  $E_x$ by the following empirical formula \cite{Ha}:
\begin{equation} \label{E}
E_x = \alpha A^{-\gamma} + \beta \left[ \exp ( - \lambda N_p ) +
\exp ( - \lambda N_n ) \right]
\end{equation}
where the first term represents the tendency toward an overall decrease of the excitation energy as the mass number $A$ increases. Since the excitation energies $E_x$ span a broad range and the differences between the measured and calculated excitation energies can be large, it is useful to introduce the logarithmic error $R_E(i)$, for the $i$-th data point, of the calculated excitation energy $E_x^{\rm cal}(i)$ with respect to the experimentally measured excitation energy $E_x^{\rm exp}(i)$ by \cite{Sabbey}
\begin{equation} \label{RE}
R_E(i) = \log \left[ E_x^{\rm cal}(i)/E_x^{\rm exp}(i) \right] =
\log \left[ E_x^{\rm cal} (i) \right] -
\log \Big[ E_x^{\rm exp} (i) \Big].
\end{equation}
Then the parameters $\alpha$, $\beta$, $\gamma$, and $\lambda$ can be fixed by minimizing the ${\chi}^2$ value which is defined by
\begin{equation} \label{Chi}
\chi^2 = { 1 \over {N_0}} \sum_{i=1}^{N_0} \left[ R_E(i) \right]^2
\end{equation}
where $N_0$ is the number of total data points considered. Note, however, that this definition is unrelated to the $\chi^2$ value usually employed in error analysis.

The apparent success of Eq.\,(\ref{E}) in reproducing the first
$2^+$ excitation energy in even-even nuclei encourages us to apply the same equation to the lowest excitation energy of other multipole states. In this work, therefore, we devote ourselves to the systematic study of an empirical expression for the lowest excitation energy of the natural parity even multipole states.

First of all, we generalize Eq.\,(\ref{E}) to the following
\begin{equation} \label{GE}
E_x = \alpha A^{-\gamma} + \beta_p \exp ( - \lambda_p N_p ) +
\beta_n \exp ( - \lambda_n N_n )
\end{equation}
in order to take into account the possibility that the contributions
to the excitation energy $E_x$ from protons and neutrons
might be different. We determine the parameters $\alpha$, $\gamma$, $\beta_p$, $\beta_n$, $\lambda_p$, and $\lambda_n$, as before,
by minimizing the $\chi^2$ value given by Eq.\,(\ref{Chi}). We perform the fitting procedure under the following four different
constraints: (I) $\beta_p=\beta_n$ and $\lambda_p=\lambda_n$, (II)
$\lambda_p=\lambda_n$, (III) $\beta_p=\beta_n$, and (IV) no
restriction on $\beta$ and $\lambda$. Case I and Case IV correspond
to Eq.\,(\ref{E}) and Eq.\,(\ref{GE}), respectively. The other two
cases correspond to the equation in between. The resulting parameter
values for each case are tabulated in Tab.\,\ref{tab-1} together
with the corresponding value of $\chi^2$. We find that the $\chi^2$
value for Eq.\,(\ref{GE}) is lower by about $17 \%$ than that for
Eq.\,(\ref{E}). Therefore, we employ Eq.\,(\ref{GE}) in
calculating the lowest excitation energy $E_x$ of the natural parity
even multipole states from now on in this work.

In each of Figs.\,1-5, we plot the excitation energy of the first
natural parity even multipole states including $2_1^+$, $4_1^+$,
$6_1^+$, $8_1^+$, and $10_1^+$, respectively, in the even-even nuclei against the mass number $A$ ($A$-plot). The upper part of these figures shows the measured excitation energies while the lower part of the same figures shows those energies calculated by our six parameter empirical formula, Eq.\,(\ref{GE}), with the parameter set as given in Tab.\,\ref{tab-2}. Observing these graphs, we see that our empirical formula can explain the essential trends of the measured lowest excitation energy of the natural parity even multipole states in even-even nuclei. By the way, the same kind of plots made by using the four parameter empirical formula, Eq.\,(\ref{E}), were shown elsewhere \cite{Yoon}. Two plots, made by using Eq.\,(\ref{GE}) and Eq.\,(\ref{E}), look almost identical although the $\chi^2$ values for the six parameter empirical formula are lower than those for the four parameter empirical formula by $17 \%$(for $2_1^+$), $14 \%$(for $4_1^+$), $8 \%$(for $6_1^+$), $4 \%$(for $8_1^+$), and $3 \%$(for $10_1^+$), respectively.

The lowest excitation energy $E_x$, given by Eq.\,(\ref{GE}), is determined by two components: one is the first term $\alpha A^{-\gamma}$ which depends only on the mass number $A$ and the other is the remaining two terms $\beta_p \exp (-\lambda_p N_p)+\beta_n \exp(-\lambda_nN_n)$ which depend only on the valence nucleon numbers, $N_p$ and $N_n$. Because there is no direct relationship between the mass number and the valence nucleon number, it would be interesting to check how the lowest excitation energy $E_x$ behaves in terms of the valence nucleon number. For that purpose, we plot the same excitation energies shown in Figs.\,1-5 again in Fig.\,\ref{fig-6} but this time against the product $N_pN_n$ ($N_pN_n$-plot). Of course, the graphs in Fig.\,\ref{fig-6} are drawn with exactly the same set of data points as those used in Figs.\,1-5. We find in Fig.\,\ref{fig-6}(a) that the measured lowest excitation energies $E_x$ show a simple pattern when the $N_pN_n$-plot is drawn. Furthermore, we also find in Fig.\,\ref{fig-6}(b) that our empirical formula reproduces the experimentally observed pattern almost exactly.
In fact, this simple pattern of the $N_pN_n$-plot was noticed a long time ago \cite{Hamamoto}. The phenomenon that a very simple pattern emerges whenever the nuclear data concerning the lowest collective state is plotted against the product $N_pN_n$ has been called the $N_pN_n$ scheme in the literature. For a long while since the idea of the $N_pN_n$ scheme was first advanced, people naively believe that the reason why the $N_pN_n$ scheme holds for the observables involving nuclear collectivity must be the active role played by the valence proton-neutron (p-n) interaction \cite{Casten1,Casten2}.

\begin{figure}[h]
\centering
\includegraphics[width=10.0cm,angle=0]{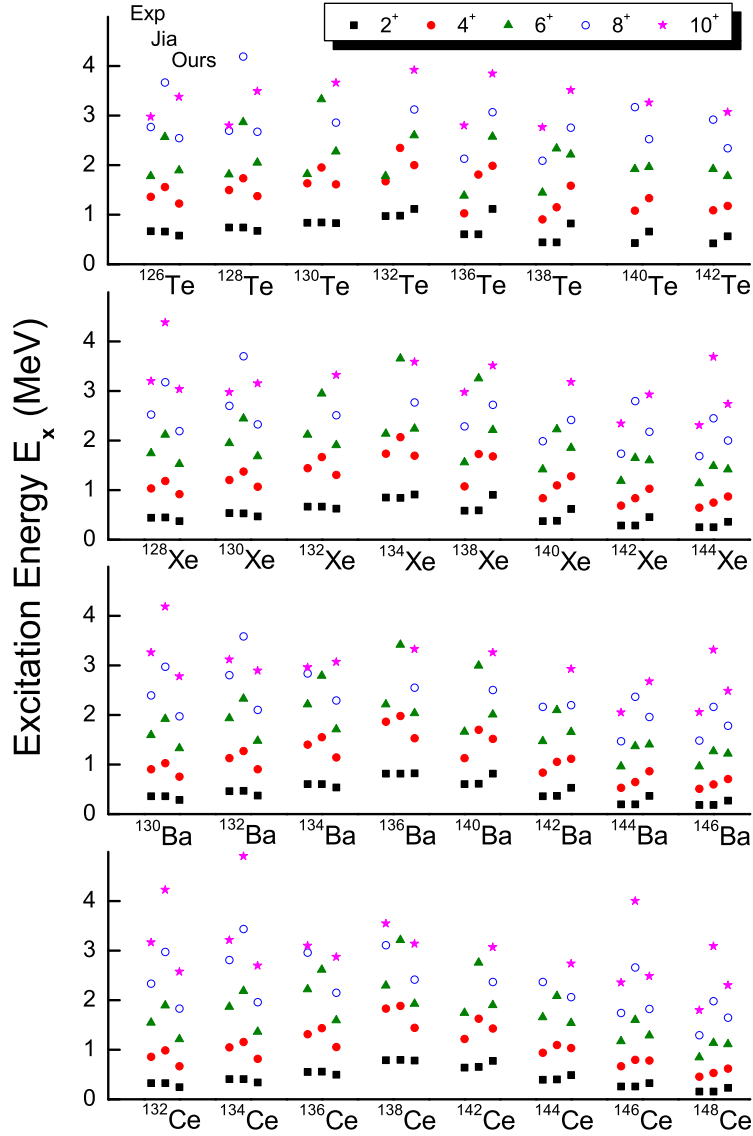}
\caption{The lowest excitation energies of the natural parity even multipole states including $2^+$, $4^+$, $6^+$, $8^+$, and $10^+$ states in even-even nuclei. The three columns marked by Exp, Jia, and Ours represent the excitation energies obtained by the experiment \cite{Firestone}, Jia {\it et al}. \cite{Jia}, and our empirical formula, respectively.}
\label{fig-7}
\end{figure}

Recently, Jia {\it et al}. published their results on the excitation energies of the low-lying states of 48 nuclides including the even-even Sn, Te, Ba, and Ce isotopes by applying the $S$ and $D$ nucleon pair approximation \cite{Jia}. We can also predict such excitation energies by our empirical formula. In Fig.\,\ref{fig-7} the two results, those by Jia {\it et al}. (the central column marked by ``Jia") and ours (the right column marked by ``Ours"), are plotted together with the measured data (the left column marked by ``Exp"). There, we show the energy spectra of nuclides with mass number $A$ ranging from 126 to 148 and with the neutron number $N$ ranging from 74 to 90. In this figure, we draw the lowest excitation energy of $2^+$ (solid squares), $4^+$ (solid circles), $6^+$ (solid triangles), $8^+$ (empty circles), and $10^+$ (solid stars) states. We can also find, from this figure, that the degree of prediction by our empirical formula is compatible with that from Jia {\it et al}.'s work. In order to compare the overall performance of the predictability between Jia {\it et al}.'s work and ours, we present the $\chi^2$ values which are calculated by Eq.\,(\ref{Chi}), the average $\bar R$ and the dispersion $\sigma$ of the logarithmic error $R_E$ in Tab.\,\ref{tab-3} where the second row marked by ``Jia" and the third row marked by ``Ours" represent Jia {\it et al}.'s work and our results, respectively. Since only $S$ and $D$ nucleon pairs are considered in Jia {\it et al}.'s work, their predictability for the excitation energy becomes worse as the multipolarity of the state increased as can be seen from Tab.\,\ref{tab-3}. On the other hand, the $\chi^2$ values obtained by our results are similar for all the multipole states considered. At any rate the overall performances of the two studies are about the same.

\begin{figure}[h]
\centering
\includegraphics[width=10.0cm,angle=0]{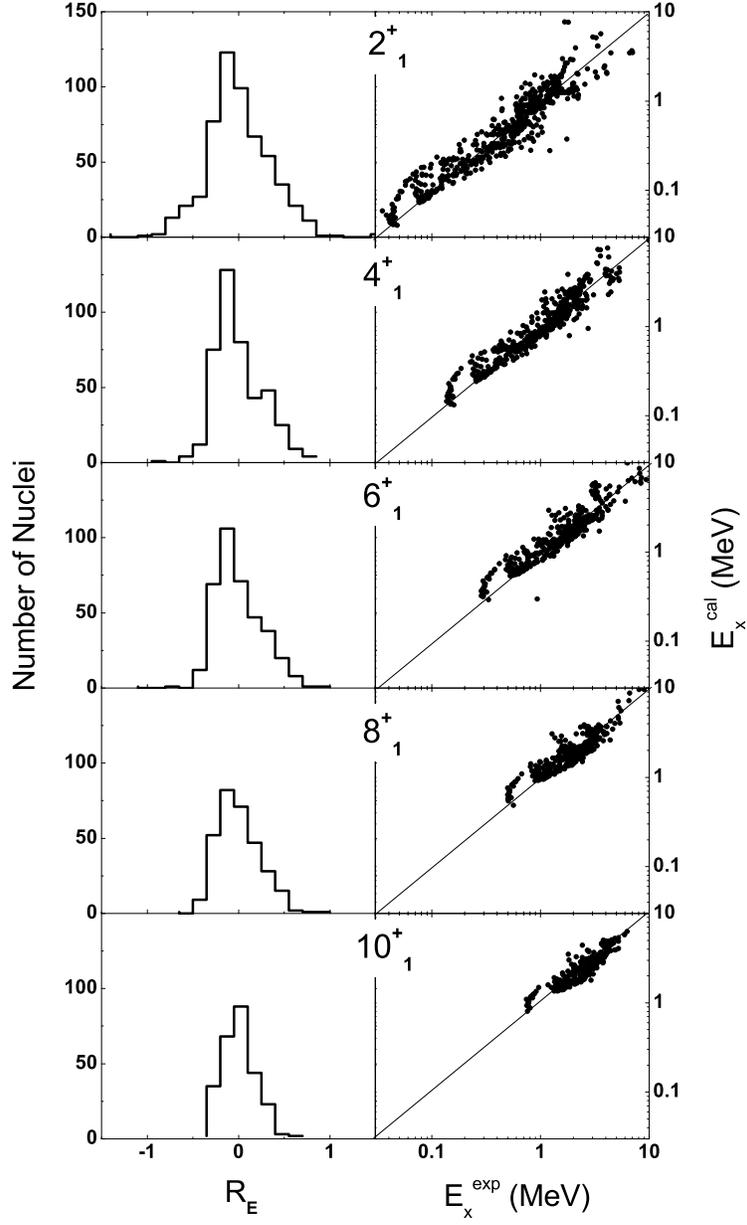}
\caption{The histogram of the logarithmic error $R_E$ against the mass number $A$ (left panels) and the scatter plot of the calculated excitation energies $E_x^{\rm cal}$ as a function of the measured ones $E_x^{\rm exp}$ (right panels) for the lowest excitation energy of the natural parity even multipole states.}
\label{fig-8}
\end{figure}

Finally, we inspect the performance of our six parameter empirical formula by drawing, in Fig.\,\ref{fig-8}, the histogram of the logarithmic error $R_E$ defined by Eq.\,(\ref{RE}) against the mass number $A$ (left panels) and the scatter plot of the calculated excitation energies $E_x^{\rm cal}$ as a function of the measured ones $E_x^{\rm exp}$ (right panels) for the lowest excitation energy of the natural parity even multipole states. Also, we show in Tab.\,\ref{tab-4}, the average $\bar R$ (the second row) and the dispersion $\sigma$ (the third row) of the logarithmic error $R_E$. From the scatter plot of Fig.\,\ref{fig-8}, we find that the number of data points which overestimate (above the line $E_x^{\rm cal}=E_x^{\rm exp}$) is about the same as the number of those which underestimate (below the line $E_x^{\rm cal}=E_x^{\rm exp}$). This is again supported by the fact that the average $\bar R$ of the logarithmic error $R_E$, as can be seen in Tab.\,\ref{tab-4}, is practically zero for all multipole states considered. From Fig.\,\ref{fig-8} and Tab.\,\ref{tab-4}, we find that our six parameter empirical formula, Eq.\,(\ref{GE}), behaves reasonably well.

In summary, we have presented an empirical formula that can be used to describe the lowest excitation energy of all of the natural parity even multipole states in even-even nuclei throughout the entire periodic table. This formula with six parameters is extended from the similar four parameter empirical formula recently introduced in our previous publication \cite{Ha}. Our empirical formula, given by Eq.\,(\ref{GE}), is composed of only three terms that depend on the mass number $A$, the valence proton number $N_p$, and the valence neutron number $N_n$, respectively. We find that Eq.\,(\ref{GE}) can explain the essential trends of the $A$-plot of the measured excitation energies as well as reproduce almost exactly the characteristic simple pattern observed in the $N_pN_n$-plot of the same measured excitation energies. We have also found that our results for the lowest excitation energy of the 48 nuclides including the even-even Sn, Te, Ba, and Ce isotopes are quite compatible with the results obtained by applying the $S$ and $D$ nucleon pair approximation by Jia {\it et al}. \cite{Jia}. In addition, we have calculated the average $\bar R$ and the dispersion $\sigma$ of the logarithmic error $R_E$ to find that our six parameter empirical formula behaves reasonably well.

\begin{acknowledgments}
We are grateful to Professor Y. M. Zhao for his valuable communications on the results of Ref. \cite{Jia}. This work was supported by an Inha University research grant.
\end{acknowledgments}

\newpage

\noindent {\bf Tables}
\\\\
\begin{table}[h]
\begin{center}
\caption{The values for the six parameters in Eq.\,(\ref{GE}) for the excitation energy of the first $2^+$ state determined by minimizing
$\chi^2$ value defined by Eq.\,(\ref{Chi}) under the following constraints: (I) $\beta_p=\beta_n$ and $\lambda_p=\lambda_n$, (II)
$\lambda_p=\lambda_n$, (III) $\beta_p=\beta_n$, and (IV) no restriction on $\beta$ and $\lambda$. In our $\chi^2$ fitting procedure, 557 measured excitation energies are used which are quoted from Ref. \cite{Raman1}.}
\begin{tabular}{cccccccc}
\hline\hline
~~~~~~&~~~~~$\alpha$~~~~~&~~~$\gamma$~~~&~~~$\beta_p$~~~&~~~$\beta_n$
~~~&~~~$\lambda_p$~~~&~~~$\lambda_n$~~~&~~~~$\chi^2$~~~~\\
&(MeV)&&(MeV)&(MeV)&&&\\
\hline
(I)&81.39&1.38&0.96&0.96&0.34&0.34&0.151\\
(II)&73.20&1.36&0.72&1.30&0.33&0.33&0.132\\
(III)&70.25&1.35&1.00&1.00&0.47&0.26&0.130\\
(IV)&68.37&1.34&0.83&1.17&0.42&0.28&0.126\\
\hline \hline
\end{tabular}
\label{tab-1}
\end{center}
\end{table}
\\\\
\begin{table}[h]
\begin{center}
\caption{The values adopted for the six parameters in Eq.\,(\ref{GE}) for the excitation energy of the first natural parity even multipole states including $2_1^+$, $4_1^+$, $6_1^+$, $8_1^+$, and $10_1^+$ states. The last two columns are the $\chi^2$ value which fits the parameter set and the total number $N_0$ of the data points, respectively, for the corresponding multipole state.}
\begin{tabular}{ccccccccc}
\hline\hline
$J_1^\pi$~~~&~~~~~$\alpha$~~~~~&~~~$\gamma$~~~&~~~$\beta_p$~~~&
~~~$\beta_n$~~~&~~~$\lambda_p$~~~&~~~$\lambda_n$~~~&~~~$\chi^2$~~~
&~~~$N_0$~~~~\\
&(MeV)&&(MeV)&(MeV)&&&&\\
\hline
$2_1^+$&68.37&1.34&0.83&1.17&0.42&0.28&0.126&557\\
$4_1^+$&268.04&1.38&1.21&1.68&0.33&0.23&0.071&430\\
$6_1^+$&598.17&1.38&1.40&1.64&0.31&0.18&0.069&375\\
$8_1^+$&1,438.59&1.45&1.34&1.50&0.26&0.15&0.053&309\\
$10_1^+$&2316.85&1.47&1.36&1.65&0.21&0.14&0.034&265\\
\hline \hline
\end{tabular}
\label{tab-2}
\end{center}
\end{table}

\newpage

\begin{table}[h]
\begin{center}
\caption{The $\chi^2$ values, the average $\bar R$, and the dispersion $\sigma$ of the logarithmic error $R_E$ calculated for the excitation energy of each multipole state which are plotted in Fig.\,\ref{fig-7}. The second row marked by Jia and the third row marked by Ours represent the $\chi^2$ values obtained by Jia {\it et al}.'s work and by our empirical formula, respectively.}
\begin{tabular}{cccccccc}
\hline\hline
\multicolumn{1}{c}~~~{$J_1^\pi$}~~~&~~~~~~~~~~&~~~~$2_1^+$~~~&~~~~$4_1^+$~~~&~~~~$6_1^+$~~~&~~~~
$8_1^+$~~~&~~~$10_1^+$~~~&~~~Total~~~\\ \hline
   &$\chi^2$&0.000&0.051&0.143&0.124&0.182&0.079\\
Jia&$\bar R$&0.000&0.139&0.349&0.338&0.413&0.200\\
   &$\sigma$&0.000&0.131&0.146&0.098&0.106&0.110 \\ \hline
   &$\chi^2$&0.098&0.086&0.066&0.049&0.026&0.070\\
Ours&$\bar R$&0.139&0.092&0.070&-0.009&0.034&0.073\\
   &$\sigma$&0.280&0.276&0.248&0.221&0.157&0.249 \\ \hline\hline
\end{tabular}
\label{tab-3}
\end{center}
\end{table}

\begin{table}[h]
\begin{center}
\caption{The average $\bar R$ and dispersion $\sigma$ of the logarithmic error $R_E$ for the lowest excitation energy of the natural parity even multipole states.}
\begin{tabular}{cccccc}
\hline\hline
~~~~~$J_1^\pi$~~~~~~&~~~~$2_1^+$~~~&~~~~$4_1^+$~~~&~~~~$6_1^+$~~~&~~~~
$8_1^+$~~~&~~~$10_1^+$~~~\\
\hline
${\bar R} \times 10^5$&-32&-83&-7&-591&152\\
$\sigma$&0.353&0.265&0.260&0.227&0.183\\
\hline \hline
\end{tabular}
\label{tab-4}
\end{center}
\end{table}

\newpage
\noindent {\bf Figures}
\\\\

\newpage

\begin{figure}[h]
\centering
\includegraphics[width=14.0cm,angle=0]{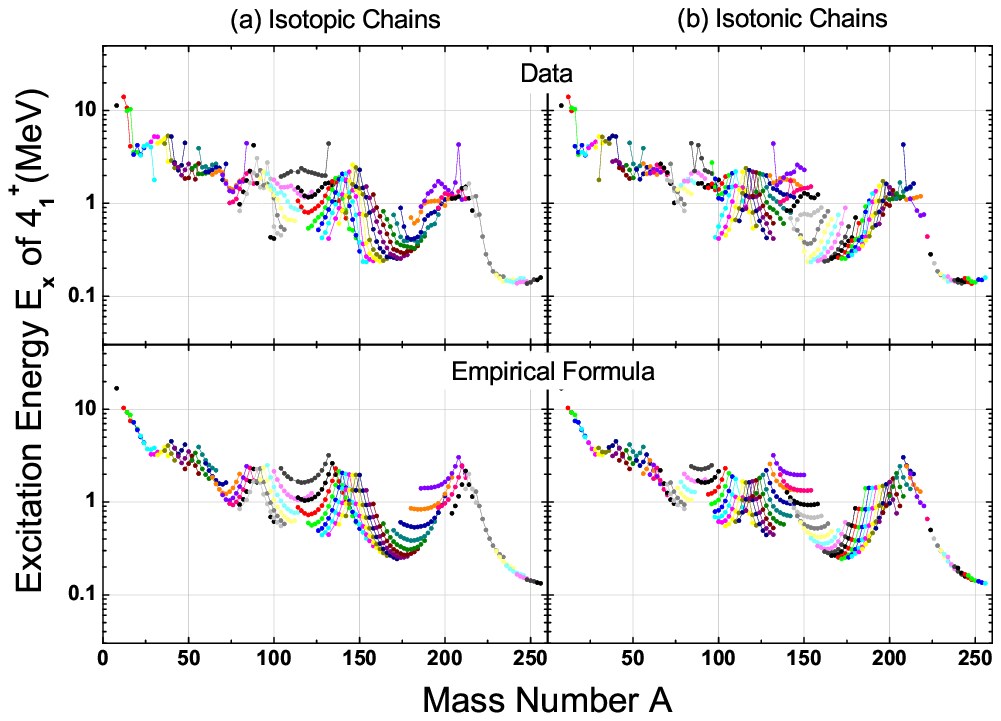}
\caption{Same as in Fig.\,1, but for the excitation energies of the first $4^+$ states in even-even nuclei. The measured excitation energies are extracted from the Table of Isotopes, 8th-edition by Firestone {\it et al}. \cite{Firestone}.}
\label{fig-2}
\end{figure}

\newpage

\begin{figure}[h]
\centering
\includegraphics[width=14.0cm,angle=0]{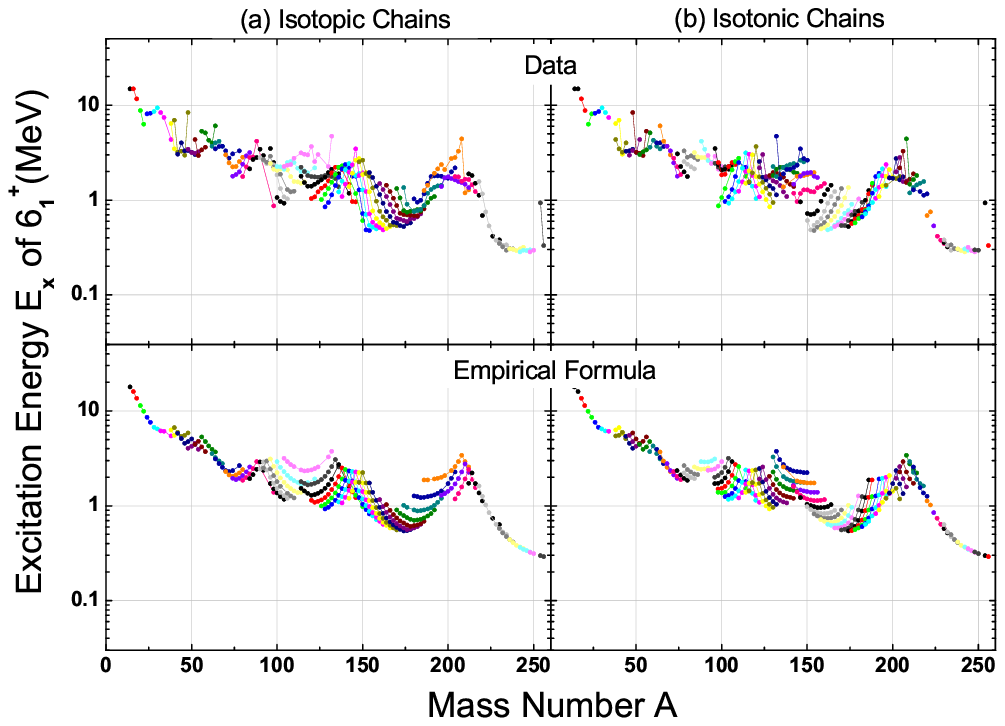}
\caption{Same as in Fig.\,1, but for the excitation energies of the first $6^+$ states in even-even nuclei. The measured excitation energies are extracted from the Table of Isotopes, 8th-edition by Firestone {\it et al}. \cite{Firestone}.}
\label{fig-3}
\end{figure}

\newpage

\begin{figure}[h]
\centering
\includegraphics[width=14.0cm,angle=0]{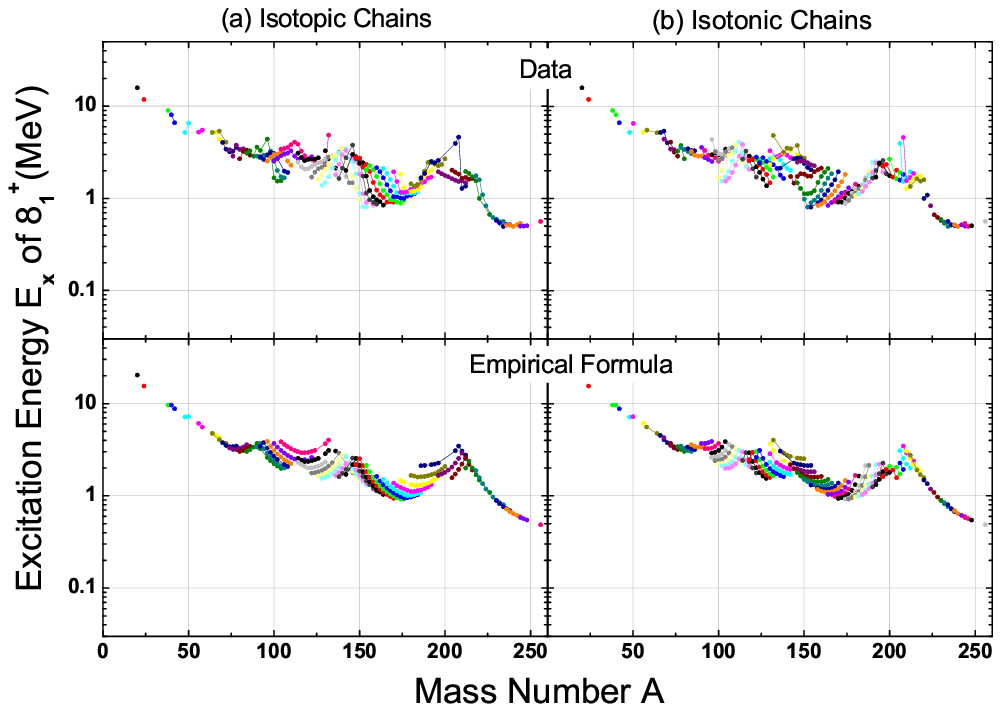}
\caption{Same as in Fig.\,1, but for the excitation energies of the first $8^+$ states in even-even nuclei. The measured excitation energies are extracted from the Table of Isotopes, 8th-edition by Firestone {\it et al}. \cite{Firestone}.}
\label{fig-4}
\end{figure}

\newpage

\begin{figure}[h]
\centering
\includegraphics[width=14.0cm,angle=0]{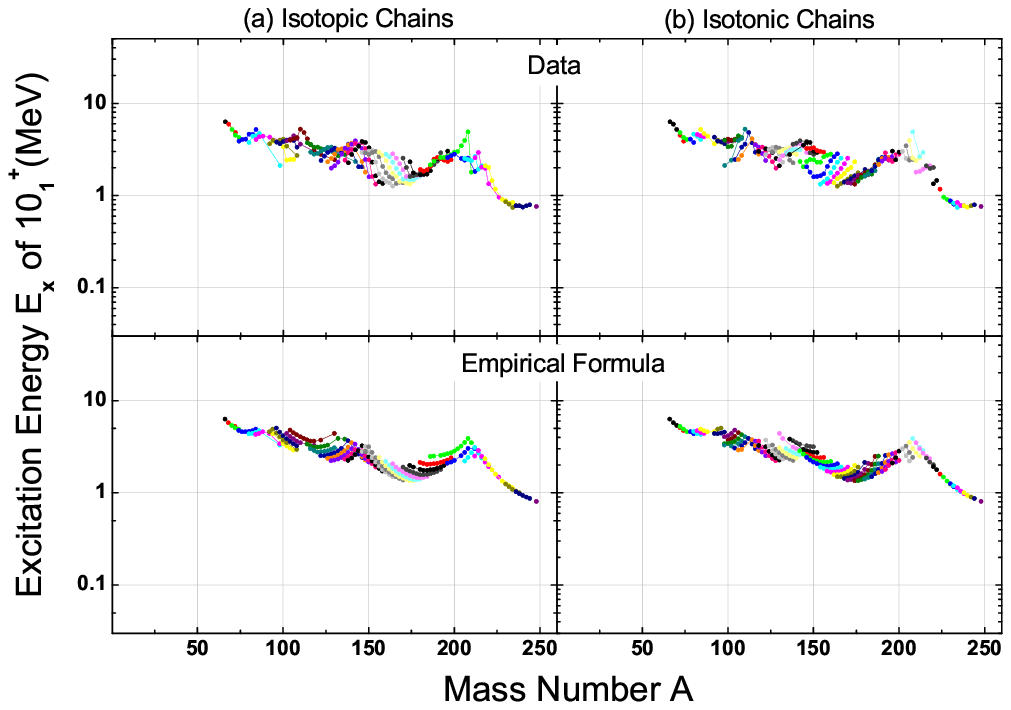}
\caption{Same as in Fig.\,1, but for the excitation energies of the first $10^+$ states in even-even nuclei. The measured excitation energies are extracted from the Table of Isotopes, 8th-edition by Firestone {\it et al}. \cite{Firestone}.}
\label{fig-5}
\end{figure}

\newpage

\begin{figure}[h]
\centering
\includegraphics[width=14.0cm,angle=0]{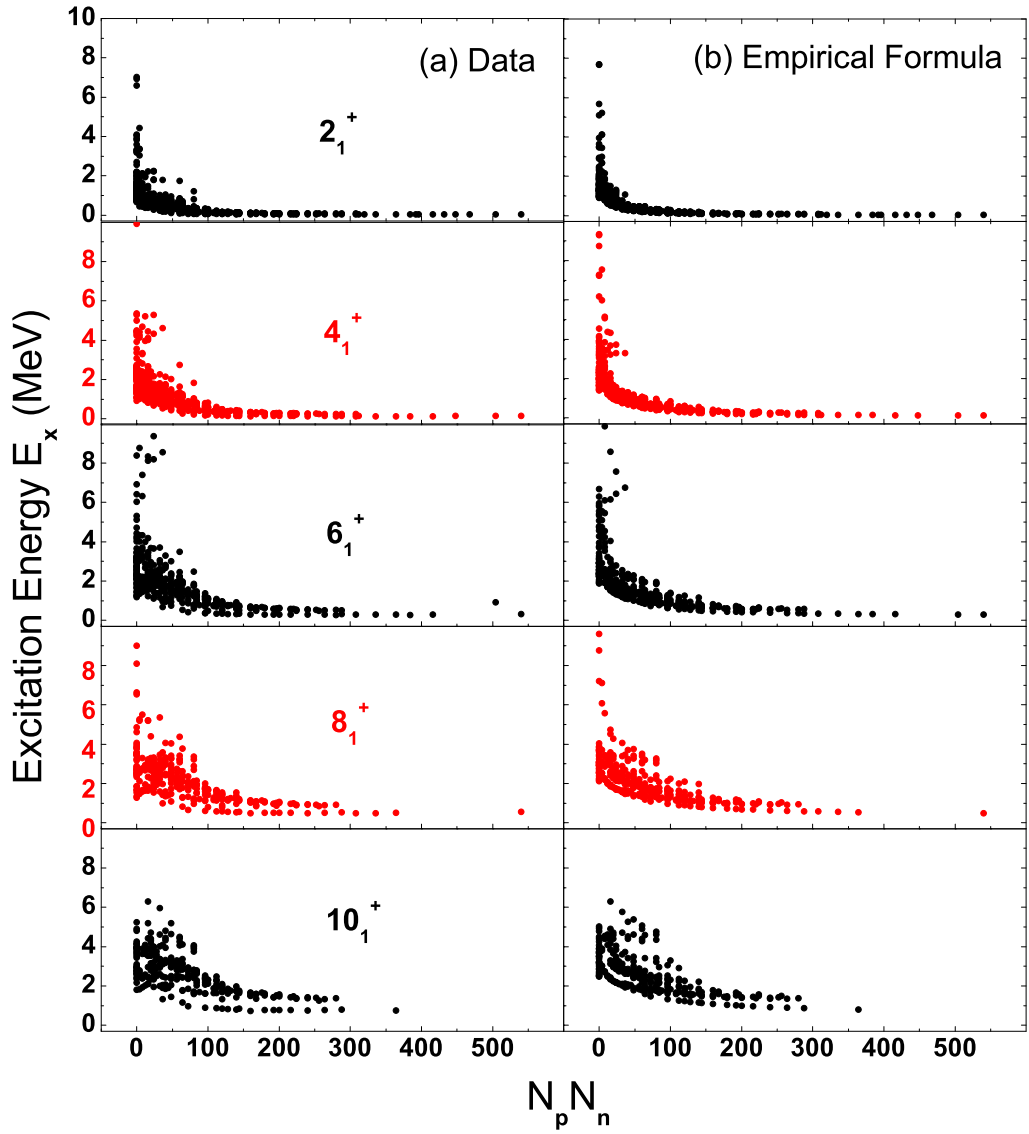}
\caption{Same as Figs.\,\ref{fig-1}-\ref{fig-5} but plotted against the product $N_pN_n$ instead of the mass number $A$.}
\label{fig-6}
\end{figure}

\newpage

\newpage

\end{document}